# Unveiling Long-Range Forces in Light Harvesting Proteins: Pivotal Roles of Temperature and Light

Elsa Perez-Martin,[1] Tristan Beranger,[1] Laurent Bonnet,[2] Frederic Teppe,[2] Alvydas Lisauskas,[3] Ketsukis Ikamas,[3] Elwin Vrouwe,[4] Elena Floriani,[5] Gergely Katona,[6] Didier Marguet,[7] Vania Calandrini,[8] Marco Pettini,[5,9] Sandra Ruffenach,[2] and Jeremie Torres[1,2]*

**Affiliations:**

[1] Institut d'Electronique et des Systèmes, University of Montpellier CNRS, Montpellier, France
[2] Laboratoire Charles Coulomb, University of Montpellier CNRS, Montpellier, France
[3] Institute of Applied Electrodynamics and Telecommunications, Vilnius University, Vilnius, Lithuania
[4] Micronit BV, Colosseum 15 PV, Enschede, Netherlands
[5] Centre de Physique Théorique, UMR 7332, Aix-Marseille Univ, Université de Toulon, CNRS, France
[6] Department of Chemistry and Molecular Biology, University of Gothenburg, 40530 Göteborg, Sweden
[7] Centre d'Immunologie de Marseille Luminy, University of Aix-Marseille CNRS INSERM, Marseille, France
[8] Computational Biomedicine (IAS-5/INM-9), Institute for Advanced Simulation, Forschungszentrum Jülich, 52428 Jülich, Germany
[9] Quantum Biology Lab, Howard University, 16 17 Washington D.C. 20059, USA

*Corresponding author: jeremie.torres@umontpellier.fr

**Abstract:** Electrodynamic interactions between biomolecules are of potential biological interest for signaling warranting investigation of their activation through various mechanisms in living systems. Here, using as model system a light harvesting protein within the phycobilisome antenna system of red algae, we proved that not only light exposure but also thermal energy alone can trigger attractive electrodynamic interactions up to hundreds of nanometer. The latter are sustained by low frequency collective modes and while the second mode appears only upon illumination, the fundamental one can be activated by temperature alone. Activation of such collective modes and ED interactions might influence conformational rearrangements and energy transport within the phycobilisome antenna system. This is a paradigm-shift that underscores the immense potential of biological systems in exploiting different forms of input energy to achieve optimal energy transfer.

# Introduction

Biological processes are highly coordinated both in time and space. Signaling events originate at the cell membrane and propagate through the cytosol (*1*). To perform their precise functions, molecular partners distributed unevenly throughout the cell must transiently interact within macromolecular ensembles (*2*, *3*). These interactions culminate in the localization of receptors, transducers, and signaling effectors within cellular compartments. At a molecular level, partners' recognition is typically accompanied by conformational changes. The interactions among them can result in the formation of a specific quaternary structure composed of multiple protein subunits, which can reversibly assemble into larger complexes. While conventional (quasi-) electrostatic interactions are crucial (*4*) for short-distance interactions, their effectiveness is limited by both Debye screening in the crowded cell environment and high permittivity values of water.

In 1968, H. Fröhlich proposed that, in out-of-equilibrium conditions, selective long-range electrodynamic (ED) forces between biomolecules can be activated when an external energy above a certain threshold is supplied (*5*). Their activation might have a strong impact on the signaling, being ED forces effective over much longer distances than electrostatic interactions (*6*). According to Fröhlich's hypothesis, such ED interactions would be sustained by giant oscillating electric dipoles, induced by a non-trivial energy condensation into (sub-TeraHertz) collective modes. Experimental evidence for energy condensation into low frequency collective modes at room temperature in open thermodynamic systems, where the energy is continuously transferred between biomolecules and their surroundings, has only recently been demonstrated (*7*, *8*). Additionally, new resonant dipole-dipole intermolecular ED forces which act over long distances of several hundreds of angstroms and directly steaming from energy condensation, have been discovered (*9*). Consequently, there has been renewed interest in the field, either from a quantum mechanical point of view (*10–13*), with the terminology of phonon condensation, or from a (semi-)classical perspective (*14–16*), in terms of collective oscillations/excitation, attracting a wide scientific audience spanning from (bio-) physics (*17*) and technology (*18*, *19*) to biology (*20*, *21*), chemistry (*22*), and even marketing (*23*).

Here using as model system R-Phycoerythrin (R-PE), a light harvesting protein naturally equipped with fluorochromes sensitive to visible light (*24*), we have compared the sub-TeraHertz (THz) spectrum of R-PE in solution upon illumination with a 488-nm laser, with the one obtained without laser, both at increasing powers of the THz probe. To compensate for the thermal energy loss upon switching off the laser, the temperature in the second protocol has been adjusted in such a way as to match the temperatures achieved in the experiment with laser. The corresponding temperature ramp varies in the interval 20 - 28 °C, which is below the protein denaturation temperature (*25*). Interestingly, while both experimental protocols feature a collective vibration at 73 GHz, a second collective vibration at 103 GHz is excited only upon laser illumination, after a saturation effect on the first peak. Because of the low frequency of the second mode, one could expect that it is always populated within the investigated temperature range, which is not the case. This suggests a non-trivial role of the electronic excitation of the protein fluorochromes in providing the protein with an additional path for energy transfer/dissipation at low frequency (103 GHz). Finally, the observed frequency dependence of those modes on protein concentration proves that they are accompanied by the activation of ED forces between proteins over distances of several hundreds of angstroms.

## Results

**Evidencing thermal activation of biomolecule collective oscillations**

Under laser light illumination conditions (see Materials and Methods), when the laser power exceeds a threshold value, R-PE diluted in a Phosphate-Buffered Saline (PBS) solution enters into a coherent vibrational state, exhibiting two distinct spectral resonances, as indicated by arrows in figure 1(a) (pink curve). These two resonances correspond to the fundamental mode (~73 GHz) and the second order mode (~103 GHz) of the collective oscillations of the R-PE protein, respectively. The frequencies of each of these modes are in agreement with previous experiments carried-out on the same living system with a different experimental setup (*9*). The high sensitivity and reliability of our novel THz-biosensor enable precise temperature control throughout experimental procedures (see Material and Methods), thus allowing the extraction of very faint THz signals from noise (Fig. S1 (*26*)). In order to verify whether the excitation of such low frequency vibrations can be explained in terms of simple thermal effects induced by the laser and the THz source (operated respectively at 60 and 100 mW and resulting in a measured temperature of 28 °C in the sample), we ran an experiment without laser but fixing the temperature at 28 °C (purple curve in figure 1(a)). Notably, only the fundamental oscillation mode is excited, whereas the second-order mode is not sustained by thermal excitation. To further investigate the interplay between optical and thermal excitations we run two sets of experiments, with and without laser, each at different THz source powers. The temperature of the setup without the laser is set in such a way as to match the corresponding temperature of the experiment with the laser. Figure 1(b) shows the corresponding normalized transmission spectra for optical excitation (solid lines) and thermal excitation (dashed lines) at different THz radiation powers. This systematic investigation further corroborates the fact that while the fundamental mode at 73

GHz is excited in both experimental protocols, the second mode at 103 GHz can be activated only in the presence of light excitation. With increasing the THz power, which results in an effective temperature increase, the mode at 103 GHz is progressively populated, after a saturation effect on the mode at 73 GHz for THz source power above 50 mW. In the experiment without light, an analogous increase of the THz power accompanied by temperature adjustment, to compensate for the loss of the laser thermal energy, does not result in the activation of the second collective vibration. Taken together, these experiments suggest that the excitation of the electronic degrees of freedom of the fluorochromes by laser light plays a fundamental role in making possible the excitation of the second collective mode at 103 GHz at room temperature. Upon light excitation, and additional efficient path for energy transfer into the low frequency collective modes is available to the system.

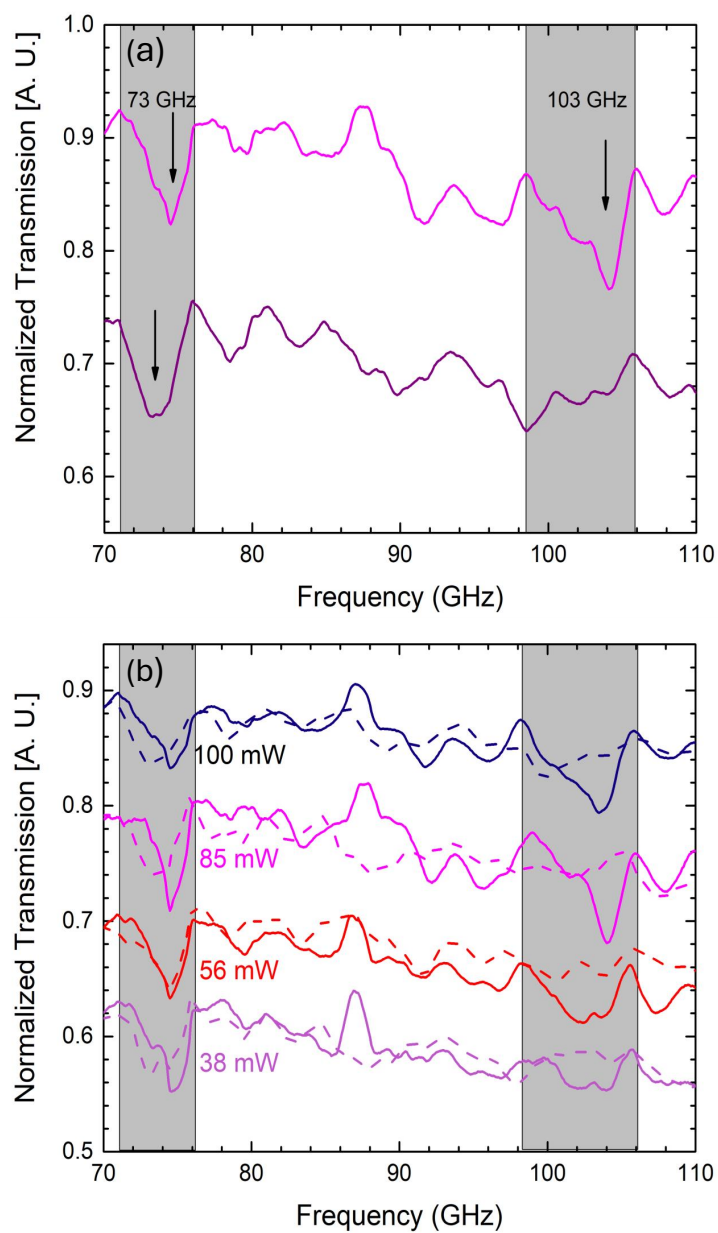

Figure 1: **Fingerprints of protein collective oscillations.** (a) Protein solutions are excited by either laser-light (pink) or thermal energy (purple). In both setups protein concentration, C, is 7.2 μM and the THz source power is 100 mW. The laser power is 60 mW. The temperature in the thermal setup is fixed to 28 °C, which matches the one achieved upon laser illumination (see text for more details). Collective oscillations at 73 GHz and 103 GHz are indicated by arrows and highlighted in gray. (b) Normalized transmission spectra upon laser illumination (continuous line) or without laser (dashed line) with varying the THz-source power from 38 to 100 mW, all at C = 4.0 μM. Spectra without illumination are taken at temperatures matching those of the setup with the laser on (see text for more details). All spectra are vertically translated for clarity.

**The activation of collective intramolecular vibrations**

Once established that the fundamental mode at 73 GHz can be thermally excited, the dependency of this collective vibration on temperature has been systematically studied in the range 20 °C to 29 °C, for three different concentrations (figure 2, panels (a), (c) and (e)). Temperature was varied following a ramp with steps of 0.1°C (see temperature control Fig. S2). To ensure consistency, PBS-buffer spectra at identical temperatures are used as references. The fundamental mode at 73 GHz is more and more populated as temperature rises above a threshold value, until becoming fully excited, with its amplitude remaining constant upon further temperature increase. The temperature interval where this transition occurs matches the optimum temperature range for red-algae growth (*25*) and is very far below the thermal unfolding of R-PE protein (*27*). Notably, for all the investigated concentrations, the variation of the quality factor $Q$ indicates that the saturation of the fundamental mode follows a threshold-like behavior (figure 2, panels (b), (d) and (f)). This feature resembles a non-equilibrium phase transition; when the energy input rate exceeds a critical value, thermal energy is channeled by the system into this coherent oscillation. Even if this behavior has been previously reported for proteins subjected to light excitation (*9*), this phase transition-like induced by the increase in temperature alone has never been reported before.

For all the investigated protein concentrations, the transition from unexcited to fully excited collective oscillation occurs in ~ 4 °C, i.e. with an extra-energy above the threshold temperature of ~ 0.3 meV (see also Fig. S4 including data at 8.0 µM). Notably, the threshold temperature decreases slightly with increasing the concentration, and the $Q$-factor values in the concentration range 1 to 7.2 µM are almost half of the one at 8.0 µM. Drawing an analogy with (collective) spontaneous emission (Schawlow – Townes law) or super-radiance mechanism (*28–30*), the $Q$-factor increase with increasing concentration is consistent with the picture that the number of proteins involved in the in-phase collective oscillation also increases. The observation that at lower protein concentration a higher energy input is required to initiate the collective oscillation suggests a possible cooperative effect between proteins.

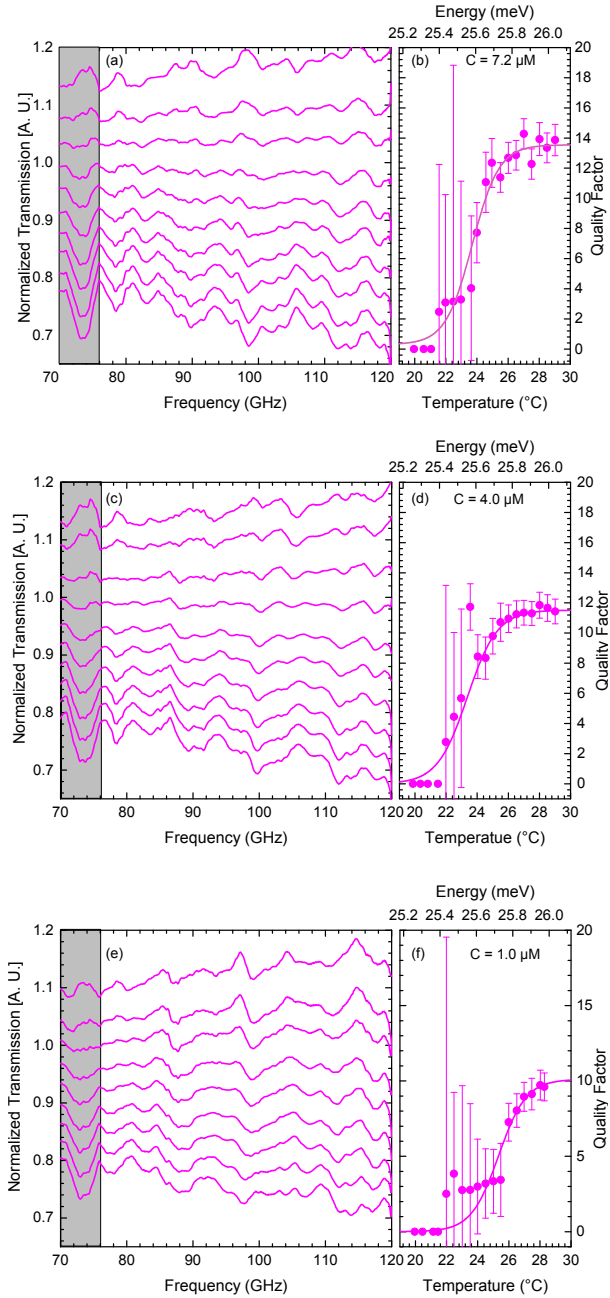

Figure 2: **Thermal activation of collective oscillations.** In all panels, temperature varies from 20 °C to 29 °C, while laser is off and THz-source power is set to 100mW. At C = 7.2 μM: (a) THz transmission spectra of R-PE normalized to PBS buffer varying the bath temperature (only spectra at intervals of 1°C are shown for clarity). (b) Fundamental mode *Q*-factor dependence with temperature. The threshold-like behavior characteristic of a phase transition is highlighted by the continuous line. At C = 4.0 μM: (c)-(d) and at C = 1.0 μM: (e)-(f). All spectra are vertically translated for clarity.

**Thermally activated long range ED interaction**

Figure 3(a) shows the normalized transmission THz spectra for different concentrations, subjected to optical (solid lines) or thermal (dashed lines) excitation. For a direct comparison, the thermal excitation spectra are provided at temperatures equivalent to those produced by laser light excitation. As previously remarked, the fundamental mode at ~73 GHz can be excited in both cases, while the second mode at ~103 GHz is excited only upon laser light illumination (starting from a protein concentration of 1.0 µM). The slight discrepancy in the fundamental mode frequency between the two excitation processes can be attributed to possibly different states explored by the proteins during these excitation processes. Figure 3(b) also depicts the normalized transmission THz spectra at different concentrations as obtained solely through thermal excitation at a fixed temperature of 28 °C. After extracting the dependency of the fundamental mode frequency $v$ on concentration, we report the corresponding normalized frequency shift variation $\Delta v = v - v_0$ versus the average intermolecular distance; $v_0$ being the unperturbed frequency at infinite dilution, i.e. $<r> \rightarrow +\infty$ (Fig. 3(c)). Remarkably, the fundamental mode frequency shifts, both for optical (blue circles) and thermal (black squares) excitation, are found to be inversely proportional to the cube of the intermolecular average distance. This serves as a proof-of-concept for the thermal activation of long-range ED attractive and selective forces, which is consistent with our classical version of the Fröhlich hypothesis (*14*). In this scenario, the proteins are all in a collective oscillation state (see also Fig. S6), leading to the activation of long-range ED forces over distances up to $<r> \sim 1350$ Å, upon thermal activation, and ~1750 Å, upon laser light excitation. It is thus clear that light is more efficient at activating intermolecular forces because the damping of the fundamental collective oscillation within this excitation mechanism is lower than in the thermal one. Experimental data have been fitted with a power law of the form $\Delta \nu/\nu_0 = A <r>^{-3}$, where A is a real number. For thermal excitation, $A = 2.74 \times 10^6$ Å$^3$ while for laser excitation, $A = 6.83 \times 10^6$ Å$^3$, in good agreement with theoretical expectation (*9*). This increase in A might be related to the enhancement in the oscillation amplitude, which entails a larger oscillating dipole moment, consistent with the increase of *Q*-values with laser light excitation. Consequently, this results in stronger intermolecular ED interactions and a larger frequency shift for a given distance (i.e. concentration). The larger strength of ED interactions generated by light could also be related to protein conformational changes under optical excitation at high concentrations, as further supported by circular dichroism experiments (Table S1, Fig. S3). At low concentrations (large inter-protein distance), the peak depicting the fundamental mode shows a similar frequency in both excitation experimental protocols, while a progressive frequency shift is observed at higher concentrations, corroborating a possible additional effect induced by protein conformational changes upon light illumination. Finally, we note that also the frequency shift of the mode at ~ 103 GHz (optical setup) as a function of the average interprotein distance (figure 3(c), red triangles), shows an $<r>^{-3}$ dependence, thus suggesting a further source of ED interactions sizably active over distances up to $<r>$ ~ 1750 Å.

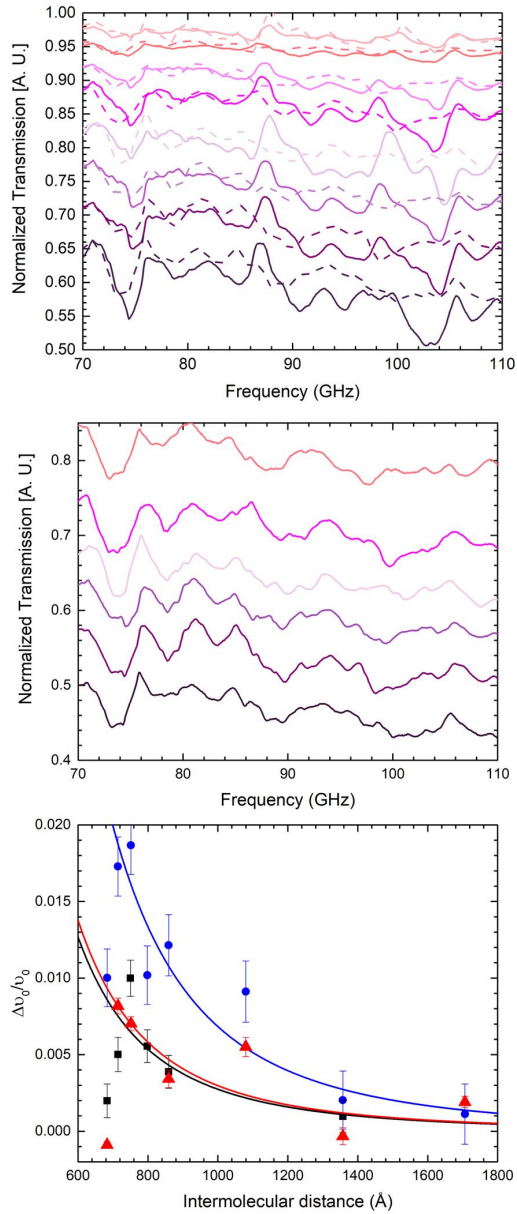

Figure 3: **Thermal activation of ED forces.** (a) Normalized transmission spectra with varying protein concentration: from top to down C = 0.5, 1.0, 2.0, 4.0, 5.0, 6.0, 7.2 and 8.0 µM. Solid lines: laser light excitation protocol (laser power = 60 mW). Dotted lines: thermal excitation protocol (laser off, temperature at each concentration is adjusted to match the temperature achieved in the corresponding laser light excitation setup. (b) Normalized transmission spectra with varying protein concentration at a fixed excitation temperature of T = 28 °C (thermal excitation protocol). Spectra are vertically translated for clarity. (c) Frequency-shift dependence to intermolecular distance for the fundamental mode upon thermal excitation at T = 28 °C (black squares), and the fundamental mode (blue circles) and the second mode upon optical excitation (red triangles). For all, THz power is set at 100 mW.

## Discussion

The reported experiments provide a proof-of-concept for the possible activation, by means of thermal energy alone, of attractive long-range electrodynamic intermolecular interactions up to ~ 1350 Å, correlated to the excitation of collective (coherent) low frequency vibrations of the protein at ~ 73 GHz. Although we could not clarify whether the excitation of such collective low frequency vibrations is accompanied by an energy condensation phenomenon as hypothesized by Fröhlich (*5*), i.e. by a redistribution of the energy from the high frequency modes to the lowest frequency one, the corresponding *Q*-factor features a typical phase transition-like evolution when temperature rises above a threshold value. Furthermore, the *Q*-factor evolution with protein concentration increase is compatible with the occurrence of in-phase collective oscillations, i.e. with the synchronization of the oscillators in the sample, which results in enhanced dipole moment fluctuations.

The mechanisms behind the activation of coherent oscillations and attractive "giant dipoles" upon excitation by "disordered" thermal energy remains elusive. Water restructurings at the water-protein interface, i.e. the dynamic exchange between free and bound water species and the strength of the hydrogen bonds (*31*), might play a role. Recently it has been proposed that sub-THz electromagnetic fields could, by reducing the orientational polarization of water molecules at the interface, decrease the dielectric permittivity of a lysozyme solution (*19*). In (*32*), anisotropic heat transfer from solvent through a protein has been discussed for soybean lipoxygenase (SLO) enzyme, a prototypical thermally activated enzyme. Specifically, it has been proved that SLO's catalytic activity involves quantum mechanical tunneling of a hydrogen atom from the substrate to the active site cofactor, which critically depends on the thermal activation of the protein scaffold. Collisions between water and a solvent exposed loop initiate thermal activation at the protein surface that is then propagated toward the reactive site through a discrete thermal network within the protein (*21*). Analogous thermal activation has been reported for thermophilic alcohol dehydrogenase (ADH) (*33*). Such anisotropic networks provide an efficient route for productive heat transfer in thermally activated enzyme reactions (*21*). The impact of thermal excitation has been investigated also in self-organized processes such as tubulin polymerization (*34*). Microtubules growth distortion upon exposure to microwave radiation, infrared laser or hot air turns out to be independent of the external excitation process as long as the corresponding thermal history is the same. Interestingly, also in R-PE both thermal and optical excitations can activate the same low frequency mode at 73 GHz when operated in equivalent temperature conditions. Conversely, the second mode at 103 GHz appears only upon laser excitation. These findings suggest that while the activation of the first vibrational mode can be related to a pure thermal effect, the second one indicates a non-trivial role of the electronic excitation of the protein fluorochromes in allowing the protein to funnel the absorbed energy into this additional low-frequency vibrational mode.

The energy conversion pathway going from the electronic excitation of the fluorochromes till the activation of such low frequency modes is an intrinsically multiscale problem both in time and space. Many papers have addressed the role of intramolecular charge transfer and vibronic excitons of the chromophores to explain energy transfer mechanisms in light harvesting proteins (see for instance (*35–38*)). Here, we provide first experimental evidence that light play a key role in the activation of a very low-frequency collective mode at 103 GHz. The corresponding frequency shift with protein concentration indicates that attractive long-range (up to ~ 1750 Å) electrodynamic interactions are in turn activated. In previous studies (*9*), we demonstrated that such electrodynamic forces, characterized by an interaction potential decreasing with the third inverse power of the intermolecular distance, lead the formation of molecular clusters. Both collective vibrations and long-range electrodynamic interactions might play a role in optimizing conformational rearrangement and energy transport within the phycobilisome antenna system, where R-Phycoerythrin (RPE) is located (*39–42*). Accurately describing such out-of-equilibrium systems, which are in constant evolution, remains a significant scientific challenge. Buffers and water might play a crucial role in sustaining collective oscillations of proteins (*43*). To address this topic a natural follow-up of this research would be the investigation of the effect of co-solvents able to modify water polarizability and viscosity. Finally, being out-of-equilibrium conditions ubiquitous in real living systems, the activation of long-range electrodynamic interactions sustained by collective vibrations could represent a general signaling mechanism, which motivates future experimental investigation of electrodynamic forces in cell.

## Acknowledgments

**Funding:** This work was supported by the Terahertz Occitanie Platform and has received funding from the European Union's Horizon 2020 Research and Innovation Programme under grant agreement # 964203 (FET-Open LINkS project) and by French national funding agency through PEPR Electronique, axis COMPTERA. A.L. and K.I. acknowledge funding received from the Lithuanian Science Foundation (project No. SMIP-22-83). **Authors contribution:** E. P-M., T.B, L.B. and S.R. have been crucial for the success carrying out all the experiments of the project both used for intramolecular collective vibration and frequency shift measurements. L.B. fixed the problem of temperature regulation while E. PM performed experiments and data treatment. L.B. and S.R. under supervision of J.T. conceived, designed, and built the experimental setup. F.T., participated in the project, supporting it since its very beginning many years ago. V.C., D.M. and G.K. gave fundamental support of biophysical and biochemical aspects and D.M. made the successful purification and time stabilization of the proteins under study. A.L. and I.K. fabricated electronic THz dice with read-out circuit. E.V. realized microfluidic cartridge design and fabrication, L.B. ensured sensor integration. M.P., E.F. and V.C. intervened in all the theoretical aspects of the project. J.T. designed, supervised, and intervened in all the experimental aspects of the project. All authors contributed to the discussion and to the analysis of the results. J.T. wrote the paper with the strong support of V.C. **Special thanks:** J.T. thanks Teraklis company (terakalis.com) for their invaluable support in designing the adaptation and holder for the Si lenses and BioCampus Montpellier platform (https://www.biocampus.cnrs.fr/index.php/fr/) for circular dichroism experiments. **Competing interests:** The authors declare that they have no competing interests. **Data and materials availability:** All data needed to evaluate the conclusions in the paper are present in the paper and/or the Supplementary Materials. All the details to reproduce the experimental outcomes and rough data are available as Supplementary Materials.